\documentclass[twocolumn]{aastex6}
\usepackage{epsfig}
\usepackage{graphicx} 
\usepackage{float}
\usepackage{amsmath}
\usepackage{color} 
\usepackage{amssymb}
\usepackage{amsfonts}
\usepackage{units}
\usepackage{bm}

\def\be{\begin{eqnarray}}
\def\ee{\end{eqnarray}}

\begin{document} 

\title{Optical Transients from Fast Radio Bursts Heating Companion Stars in Close Binary Systems}

\author{Yuan-Pei Yang\altaffilmark{1}} 

\affil{
$^1$ South-Western Institute for Astronomy Research, Yunnan University, Kunming, Yunnan, 650500, P.R.China; ypyang@ynu.edu.cn
}

\begin{abstract} 

Fast radio bursts (FRBs) are bright radio transients with short durations and extremely high brightness temperatures, and their physical origins are still unknown. Recently, a repeating source, FRB 20200120E, was found in a globular cluster in the very nearby M81 galaxy. The associated globular cluster has an age of $\sim9.13~{\rm Gyr}$, and hosts an old population of stars. In this work, we consider that an FRB source is in a close binary system with a low-mass main sequence star as its companion. Due to the large burst energy of the FRB, when the companion star stops the FRB, its surface would be heated by the radiation-induced shock, and make re-emission. For a binary system with a solar-like companion star and an orbital period of a few days, we find that the re-emission is mainly at optical band, and with delays of a few seconds after the FRB. Its luminosity is several times larger than the solar luminosity, and the duration is about hundreds of seconds. Such a transient might be observable in the future multiwavelength follow-up observation for Galactic FRB sources.

\end{abstract} 


\section{Introduction} 

Fast radio bursts (FRBs) are bright radio transients with short durations and extremely high brightness temperatures \citep[e.g.,][]{Lorimer07,Thornton13}, and their physical origins are still unknown \citep[e.g.,][]{Petroff19,Cordes19,Zhang20,Xiao21}. Up to the present, hundreds of FRB sources have been detected, and dozens of them showed repeating behaviors \citep[e.g.,][]{CHIME21}.
The recent discovery of Galactic FRB 200428 associated with SGR J1935+2154 suggests that at least some FRBs originate from magnetars born from core collapse of massive stars \citep{Bochenek20,CHIME20,Li20,Mereghetti20,Ridnaia20,Tavani20}, but the emission region and the coherent mechanism are still not confirmed \citep[e.g.,][]{Katz16,Murase16,Beloborodov17,Kumar17,Yang18,Yang21,Metzger19,Lu20,Margalit20,Wadiasingh20,Ioka20,Lyubarsky21,Wang21}. 

Very recently, a repeating FRB source, FRB 20200120E, was found to be associated with a globular cluster in the very nearby M81 galaxy at $3.6~{\rm Mpc}$ \citep{Bhardwaj21,Kirsten21}. 
Since globular clusters host old stellar populations, such an association challenges FRBs originating from active magnetars born from core-collapse supernovae, and implies that there might be more than one formation channel for magnetars as the central engines of FRBs \citep{Kremer21,Lu21}. 
Besides, a repeating source, FRB 180916B with 16-day periodic activity \citep{CHIME20b}, was localized to be $\sim 250~{\rm pc}$ offset from the brightest region of the nearest young stellar clump that was proposed to be its birth place \citep{Tendulkar21}, which means that the source of FRB 180916B would take 0.8-7 Myr to traverse to the current observed position with the typical projected velocities ($60-750~{\rm km~s^{-1}}$) of neutrons stars in binaries. 
Because the traversal timescale is inconsistent with the active times ($\lesssim10~{\rm kyr}$) of a magnetar, this observation also challenges that FRB 180916B was emitted from a magnetar born from a core-collapse supernova. Meanwhile, the observed 16-day periodic activity also implies that it may be in a binary system as proposed by some FRB models \citep{Zhang17,Zhang20d,Dai20,Zhang20c,Ioka20b,Lyutikov20,Deng21,Sridhar21,Geng21,Wada21,Li21c}

The detection of associated multiwavelength or multi-messenger counterparts is helpful to identify the physical origin and radiation mechanism of FRBs.
In general, there are four physical mechanisms that can give rise to an FRB-associated multiwavelength counterpart: 1. the extension of the radiation mechanism of FRB emission to higher frequencies \citep[e.g.,][]{Yang19}; 2. the multiwavelength afterglow associated with an FRB explosion energy \citep{Yi14,Wang20c}; 3. the inverse Compton scattering processes associated with an FRB \citep{Yang19}; 4. the astrophysical events directly associated with FRBs \citep{Beloborodov17,Metzger19}.

However, in addition to an X-ray burst associated FRB 200428 \citep{Li20,Mereghetti20,Ridnaia20,Tavani20}, there is no confirmed multiwavelength or multi-messenger transient being associated with other FRBs so far, which might be due to three main reasons \citep[e.g.,][]{Wang20d}: 1. the typical fluxes of the multiwavelength counterparts of FRBs are lower than the sensitivities of current detectors, like faint FRB afterglows \citep{Yi14}; 2. the durations of the multiwavelength transients are shorter than the time resolutions of detectors, like fast optical bursts \citep{Yang19,Tingay19}; 3. the delay time between FRB and its associated transient is much longer than the observation time, like the scenario of GRB-FRB association \citep{Wang20e,Wang20d}.

Inspired by the recent discovery of FRB 20200120E associated with a globular cluster in the M81 galaxy, we are interested in the process that an FRB heats the companion star in a close binary system and the corresponding multiwavelength radiation.
In a globular cluster similar to the host of FRB 20200120E, the numbers of main sequence stars, white dwarfs and neutron stars might be about $10^5$, $10^4$ and $10^2$, respectively \citep{Kremer21}. We assume that neutron stars are the engines of FRBs as proposed by most FRB models \citep[e.g.,][]{Kumar20,Lu20,Yang21}, and they would be preferred to have a companion as low-mass main sequence stars in the globular cluster.
Due to the large burst energy of FRB, when the companion star stops an FRB, its surface would be heated and make re-emission, as shown in Figure \ref{fig1}. We discuss the radiation from the heated companion star in Section \ref{sec2}, including the cases of the companion star not filling / filling Roche lobe in Section \ref{sec21} and Section \ref{sec22}, respectively. The results are summarized in Section \ref{sec3} with some discussions. The convention $Q_x=Q/10^x$ is adopted in cgs units, unless otherwise specified.

\section{FRB heating a companion star}\label{sec2}

\begin{figure}[]
    \centering
	\includegraphics[width = 0.9\linewidth, trim = 50 150 50 150, clip]{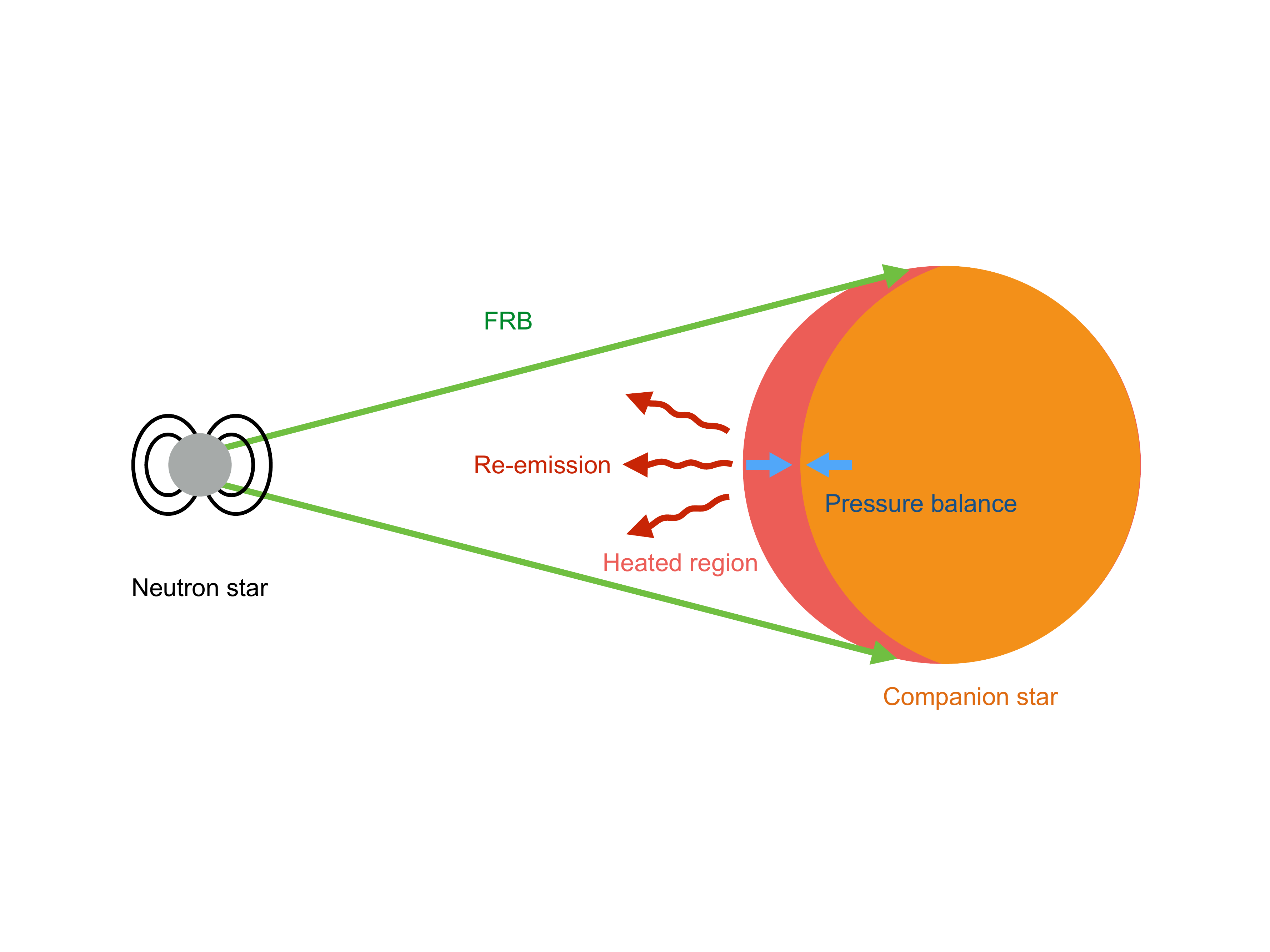}
    \caption{Schematic configuration: FRB heats the surface of the companion star in a close binary system and re-emits at optical bands.}\label{fig1}
\end{figure}

Globular clusters are believed to host Population II stars with the mass-age relation satisfying \citep{Lejeune01,Schaerer02,Cooray}
\be
\log\left(\frac{\tau_{\rm age}}{\rm yr}\right)=9.59-2.79\log\left(\frac{M}{M_\odot}\right)+0.63\left[\log\left(\frac{M}{M_\odot}\right)\right]^2.\nonumber\\
\ee
For a globular cluster with age $\tau_{\rm age}\sim 9.13~{\rm Gyr}$ \citep{Kirsten21}, like the host of FRB 20200120E, the mass of Pop II stars inside it required to be 
\be
M\lesssim0.75 M_\odot.
\ee 
Thus, in the following discussion, we focus on a close binary system with an FRB source, i.e., neutron star, and a low-mass main sequence star. We assume that the neutron star has mass of $M_{\rm NS}\sim 1.4M_\odot$, and the companion star has a possible mass range of $M=(0.1-1)M_\odot$.

According to the Kepler's third law, for a binary system with orbital period $P$ and total mass $M_{\rm tot}=M+M_{\rm NS}$, the semimajor axis of the binary is
\be
a=\left(\frac{GM_{\rm tot}}{\Omega^2}\right)^{1/3}=(3.4-3.9)\times10^{11}~{\rm cm}~P_{\rm day}^{2/3},\label{K3}
\ee
where $P_{\rm day}=P/(1~{\rm day})$ and $M_{\rm tot}=M+M_{\rm NS}\sim(1.5-2.4)M_\odot$ is taken. We take $a\sim3.65\times10^{11}~{\rm cm}~P_{\rm day}^{2/3}$ as an approximation in follows. If an FRB heats the companion star and makes re-emission in the binary system, the intrinsic time delay between the FRB and the re-emission is about
\be
t_{\rm delay}\sim\frac{a\cos i}{c}\simeq12~{\rm s}~P_{\rm day}^{2/3}\cos i,
\ee
where $i$ is the orbital inclination of the binary system, and a detailed result is dependent of the relative position of both stars. If the companion star is very close to the FRB source, a large radiation pressure of FRB would act on the surface of the companion star, 
\be
P_{\rm FRB}&\simeq&\frac{E_{\rm FRB}}{4\pi c a^2\Delta t_{\rm FRB}}\nonumber\\
&=&2\times10^7~{\rm dyne~cm^{-2}}E_{\rm FRB,39}\Delta t_{\rm FRB,-3}^{-1}P_{\rm day}^{-4/3},\label{pressure}
\ee
where $E_{\rm FRB}=10^{39}~{\rm erg}$ is the isotropic energy in radio emission emitted by an FRB. Here, we take the typical value based on the luminosities of most extragalactic FRBs \citep[e.g.,][]{Luo20}.
Some physical scenarios proposed that the total radio fluence is only a small fraction of the total energy required by the FRB central engine \citep[e.g.,][]{Metzger19,Lu20,Margalit20}, in which case the heating effect by total pressure (including radiation pressure at all bands and particle gas pressure) would be more significant than that given by Eq.(\ref{pressure}). 
However, a low radio emission fraction, e.g., $\xi\sim10^{-5}$ for FRB 200428, implies that the active age of a repeating FRB source would be very short \citep{Yang21}, which is inconsistent with the observation of FRB 121102 with the large burst event rate for many years \citep{Li21}. 
In the following discussion, at a conservative estimate, we mainly focus on the energy contribution by radio emission of an FRB.
The radiation pressure given by Eq.(\ref{pressure}) is much larger that the radiation pressure at the companion star surface
$P\simeq L/4\pi c R^2=2~{\rm dyne~cm^{-2}}L_{0,\odot}R_{0,\odot}^{-2}$, where $L_{0,\odot}=L/L_\odot$, $R_{0,\odot}=R/R_{\odot}$.
The intense radiation pressure of the FRB pulse would push an overdense target inwards, steepening the density profile, and a shock would be generated in this process, as shown in Figure \ref{fig1}.

The radiation-induced shock sweeps the surface medium of the companion star and finally be choked at the region where the stellar pressure is balanced with the radiation pressure, as shown in Figure \ref{fig1}.
Before analyzing how many the surface medium are heated by the shock, we first make a discussion about the stellar structure.
For a spherically symmetric star in hydrostatic equilibrium, the basic structure equations involve
$\partial r/\partial m=1/4\pi r^2\rho$
and
$\partial P/\partial m=-Gm/4\pi r^4$,
where $P$ is the pressure, $\rho$ is the mass density, and $m$ is the mass inside $r$. For the uniform-density stellar model\footnote{Although we take a uniform-density stellar model as an approximation here, the following result of pressure-mass relation is consistent with the standard solar model \citep[e.g.,][]{Guenther92}.}, $\rho={\rm const}$, one has $r=(3m/4\pi\rho)^{1/3}$. Eliminating $r$, one obtain from the hydrostatic equation $d P/d m=-(G/4\pi)(4\pi\rho/3)^{4/3}m^{-1/3}$, which can be integrated to yield
\be
P-P_{0}=-\frac{3G}{8\pi}\left(\frac{4\pi}{3}\rho\right)^{4/3}m^{2/3}.
\ee
At the star surface, due to $m=M$ and $P\sim0$, one obtain $P_{0}=(3G/8\pi)(4\pi\rho/3)^{4/3}M^{2/3}$. Then the mass at the region where the pressure less than $P$ can be calculated by
\be
m(<P)=M-\left[M^{2/3}-\frac{8\pi P}{3G}\left(\frac{M}{R^3}\right)^{-4/3}\right]^{3/2}\simeq\frac{4\pi P R^4}{G M}.\nonumber\\\label{mass}
\ee
The radiation-induced shock is finally choked at the region where the pressure is balanced, $P=P_{\rm FRB}$, and one has $m(<P_{\rm FRB})=(10^{-8}-10^{-10})M_{\odot}$ for a companion star with mass of $M=(0.1-1)M_\odot$. Notice that the mass given by Eq.(\ref{mass}) is isotropic, and a detailed analysis will be discussed in Section \ref{sec21} and Section \ref{sec22}.
Therefore, the larger the radiation pressure of FRBs, the larger the swept mass.

Since the FRB pulse cannot penetrate into the companion star, the absorbed energy is firstly transported mostly by energetic electrons, and further transfers to ions via collision processes. The energy of energetic electrons is of the order of the cycle-averaged oscillation energy in the electric field of the FRB in vacuum \citep{Macchi13,Yang20b},
\be
\varepsilon_e=(\gamma-1)m_ec^2=m_ec^2\left(\sqrt{1+a_{\rm str}^2/2}-1\right),
\ee
where the Lorentz factor $\gamma=(1+a_{\rm str}^2/2)^{1/2}$ of energetic electrons depends on the strength parameter with \citep[e.g.,][]{Yang20b}
\be
a_{\rm str}&=&\frac{eE}{m_ec\omega}=\frac{e L_{\rm FRB}^{1/2}}{2\pi m_ec^{3/2}\nu a}\nonumber\\
&=&44 E_{\rm FRB,39}^{1/2}\Delta t_{\rm FRB,-3}^{-1/2}\nu_9^{-1}P_{\rm day}^{-2/3},
\ee
where $E$ is the oscillating electric field of electromagnetic wave, and $L_{\rm FRB}$ is the isotropic luminosity of FRB. Therefore, the typical energy of energetic electrons is estimated by
\be
\varepsilon_e\sim\frac{a_{\rm str}}{\sqrt{2}}m_ec^2=16~{\rm MeV}~E_{\rm FRB,39}^{1/2}\Delta t_{\rm FRB,-3}^{-1/2}\nu_9^{-1}P_{\rm day}^{-2/3}.\nonumber\\
\ee
At the depth with density of $\rho\sim10^{-5}~{\rm cm^{-3}}$ where the pressure is balanced as discussed in follows (see Eq.(\ref{rho1}) and Eq.(\ref{rho2})), the typical timescale of electron-electron collisions is \citep[e.g.,][]{Somov12}
\be
t_{ee}&\sim&\frac{m_e^2(\varepsilon_e/m_e)^{3/2}}{\pi e^4(\rho/m_p)}\nonumber\\
&=&4~{\rm ms}~E_{\rm FRB,39}^{3/4}\Delta t_{\rm FRB,-3}^{-3/4}\nu_9^{-3/2}P_{\rm day}^{-1}\rho_{-5}^{-1}.
\ee
For the same typical parameters, the timescale of proton-proton collisions is $t_{pp}\simeq43 t_{ee}=0.2~{\rm s}$, and the timescale of electron-proton collisions is $t_{pe}\simeq950 t_{ee}=3.7~{\rm s}$ (see Section 8.3 of \citet{Somov12}). Therefore, at the pressure balance region, most particles would be heated and become thermal in a short time compared with the radiation timescale.
In the following discussion, we will analyze the re-emission process of an FRB heating the companion star, including two cases: 1. the companion star not filling its Roche lobe (Section \ref{sec21}); 2. the companion filling its Roche lobe (Section \ref{sec22}).

\subsection{Case A: companion star not filling Roche lobe}\label{sec21} 

First, we consider that the companion star in the binary system does not fill its Roche lobe. For an isolated star with mass $M\lesssim1~M_{\odot}$, the mass-radius relation satisfies
\be
R\simeq R_{\odot}\left(\frac{M}{M_{\odot}}\right)^{0.8}~~{\rm or}~~M\simeq M_{\odot}\left(\frac{R}{R_{\odot}}\right)^{1.25},\label{MR}
\ee
where $M_\odot=2\times10^{33}~{\rm g}$ and $R_\odot=7\times10^{10}~{\rm cm}$ \citep[e.g.,][]{Kippenhahn12}.
Thus, the low-mass main sequence stars with mass $M\sim(0.1-1)M_\odot$ have radius of $R\sim(0.16-1)R_\odot$. 
We assume that the FRB emission region is inside the binary orbit, as proposed by most ``close-in'' models \citep[e.g.,][]{Kumar20,Lu20,Yang20}, then the solid angle of the companion star opened to the FRB source is
\be
\Delta\Omega\sim \frac{\pi R^2}{a^2}=0.12R_{0,\odot}^2P_{\rm day}^{-4/3}\label{Omega}
\ee
for $R\ll a$.
We assume that the radiation beaming of FRBs, $\Delta\Omega_{\rm FRB}$, is larger than $\Delta\Omega$.
It is not clear whether the FRB emission mechanism is beamed. If the FRB radiation is highly beamed, only a part of the companion surface would be heated, meanwhile, the radio burst would fail to interact with the comapnion star for a large fraction of the time. Recently, \citet{Connor20} proposed that a selection effect due to beamed emission causes the observed difference in bursts durations of repeating FRBs and one-off FRBs. In this scenario, the relation between the beaming solid angle and duration is proposed to satisfies $\Delta\Omega_{\rm FRB}\sim0.2\Delta t_{\rm FRB,-3}$, which is slightly larger than $\Delta\Omega$ given by Eq.(\ref{Omega}).
According to Eq.(\ref{K3}) and Eq.(\ref{Omega}), the radiation energy of the FRB emission toward to the companion star is 
\be
E_{\Delta\Omega}\simeq\left(\frac{\Delta\Omega}{4\pi}\right)E_{\rm FRB}\sim10^{37}~{\rm erg}~E_{\rm FRB,39}R_{0,\odot}^2P_{\rm day}^{-4/3}.\nonumber\\\label{energy1}
\ee
The large radiation pressure would sweep the surface medium of the companion star and finally be choked at the pressure balance region. According to Eq.(\ref{pressure}), Eq.(\ref{mass}) and Eq.(\ref{MR}), the mass of shocked medium is given by
\be
m(<P_{\rm FRB})
=2.3\times10^{-8}M_{\odot}E_{\rm FRB,39}\Delta t_{\rm FRB,-3}^{-1}P_{\rm day}^{-4/3}R_{0,\odot}^{11/4}.\nonumber\\
\ee
We notice that the above mass is isotropic, and only a fraction $f_b\sim(0.1-1)$ of the surface medium within $\Delta\Omega$ could be swept by the radiation pressure of the FRB, as shown in Figure \ref{fig1}. Therefore, the number of shocked particles is given by
\be
N&\simeq&\frac{f_bm(<P_{\rm FRB})}{m_p}=2.8\times10^{48}f_{b,-1}\nonumber\\
&\times&E_{\rm FRB,39}\Delta t_{\rm FRB,-3}^{-1}P_{\rm day}^{-4/3}R_{0,\odot}^{11/4}.\label{number1}
\ee
The radiation energy would transfer to thermal energy of particles by the shock, and the temperature of the shocked medium is estimated by
\be
kT\simeq\frac{\eta E_{\Delta\Omega}}{N}=2.2~{\rm eV}\eta f_{b,-1}^{-1}\Delta t_{\rm FRB,-3}R_{0,\odot}^{-3/4},\label{temperature1}
\ee
where $\eta$ is the absorption coefficient that is contributed by free-free absorption, synchrotron absorption and plasma absorption \citep[e.g.,][]{Yang16,Kundu21}. It is noteworthy that the particle temperature is independent of the FRB burst energy. The reason is that the mass of shocked medium is proportional to the burst energy according to Eq.(\ref{pressure}) and Eq.(\ref{mass}), leading to the accelerated energy of each particle is independent of the burst energy after thermalization.
According to Eq.(\ref{pressure}) and Eq.(\ref{temperature1}), the mass density of the shocked medium is
\be
\rho&\simeq&\frac{m_p P_{\rm FRB}}{kT}=10^{-5}~{\rm g~cm^{-3}}~\eta^{-1} f_{b,-1}\nonumber\\
&\times&E_{\rm FRB,39}\Delta t_{\rm FRB,-3}^{-2}P_{\rm day}^{-4/3}R_{0,\odot}^{3/4},\label{rho1}
\ee
and the thickness of shocked medium is estimated by
\be
l\simeq\frac{f_bm(<P_{\rm FRB})}{2\pi R^2 \rho}=1.5\times10^7~{\rm cm}~\eta\Delta t_{\rm FRB,-3}.\label{l1}
\ee
Based on Eq.(\ref{rho1}) and Eq.(\ref{l1}), the optical depth for Thomson scattering is 
\be
\tau\simeq\kappa\rho l=60~ f_{b,-1}E_{\rm FRB,39}\Delta t_{\rm FRB,-3}^{-1}P_{\rm day}^{-4/3}R_{0,\odot}^{3/4},
\ee
where $\kappa\sim0.4~{\rm cm^{2}g^{-1}}$ is the opacity contributed by Thomson scattering of fully ionized hydrogen. 
One can see that the larger the separation, the smaller the optical depth.
The re-emission depends on the effective temperature for the black-body radiation. 
According to the theory of radiative transfer, the effective temperature depends on the optical depth,
\be
T_{\rm eff}=T\left(\frac{1}{2}+\frac{3}{4}\tau\right)^{-1/4}.
\ee
In the following discussion, we will discuss both cases with $\tau\gg1$ and $\tau\lesssim1$.
First, for the case with $\tau\gg1$ ($P\ll22~{\rm day}$ with the above typical parameters), the effective temperature is
\be
kT_{\rm eff}&\simeq& kT\left(\frac{3}{4}\tau\right)^{-1/4}=0.85~{\rm eV}~\eta f_{b,-1}^{-5/4}\nonumber\\
&\times&E_{\rm FRB,39}^{-1/4}\Delta t_{\rm FRB,-3}^{5/4}P_{\rm day}^{1/3}R_{0,\odot}^{-15/16}.
\ee
Considering that only about half of the surface area could be heated, as shown in Figure \ref{fig1}, the luminosity of re-emission by the FRB heating the companion is estimated as
\be
L_{\rm re}&\simeq&2\pi R^2\sigma T_{\rm eff}^4=1.6\times10^{34}~{\rm erg~s^{-1}}\eta^4 f_{b,-1}^{-5}\nonumber\\
&\times&E_{\rm FRB,39}^{-1}\Delta t_{\rm FRB,-3}^5P_{\rm day}^{4/3}R_{0,\odot}^{-7/4},\label{lum1}
\ee
which is much larger than the solar luminosity. 
The typical timescale of re-emission is about
\be
t_{\rm re}&\simeq&\frac{\eta E_{\Delta\Omega}}{L_{\rm re}}=620~{\rm s}~\eta^{-3} f_{b,-1}^{5}\nonumber\\
&\times&E_{\rm FRB,39}^2\Delta t_{\rm FRB,-3}^{-5}P_{\rm day}^{-8/3}R_{0,\odot}^{15/4}.\nonumber\\
\ee
Eq.(\ref{lum1}) gives the luminosity of the re-emission at all band.
For an optical band at $\lambda\simeq5000 \text{\AA}$, the observed flux at distance $d$ is estimated by
\be
F_\nu&=&\pi B_\nu\left(\frac{R}{d}\right)^2\simeq\frac{2\pi\nu^2}{c^2}kT_{\rm eff}\left(\frac{R}{d}\right)^2=1.8\times10^{-3}~{\rm Jy}\nonumber\\
&\times&\eta f_{b,-1}^{-5/4}E_{\rm FRB,39}^{-1/4}\Delta t_{\rm FRB,-3}^{5/4}P_{\rm day}^{1/3}R_{0,\odot}^{17/16}\lambda_{\rm opt}^{-2}d_{\rm 1,kpc}^{-2},\label{flux1a}
\ee
where $d_{\rm 1,kpc}=d/(10~{\rm kpc})$ and $\lambda_{\rm opt}=\lambda/5000 \text{\AA}$ is adopted.

On the other hand, for the case with $\tau\lesssim1$ ($P\gg22~{\rm day}$ with the above typical parameters), the effective temperature is $T_{\rm eff}=2^{1/4}T$. The luminosity of re-emission by the FRB heating the companion is
\be
L_{\rm re}&\simeq&2\pi R^2\sigma T_{\rm eff}^4\nonumber\\
&=&1.5\times10^{36}~{\rm erg~s^{-1}}\eta^4 f_{b,-1}^{-4}\Delta t_{\rm FRB,-3}^4R_{0,\odot}^{-1},\label{lum2}
\ee
which is about a few hundreds times than the solar luminosity.
The typical timescale of re-emission is about
\be
t_{\rm re}\simeq\frac{\eta E_{\Delta\Omega}}{L_{\rm re}}=6.7~{\rm s}~\eta^{-3} f_{b,-1}^{4}E_{\rm FRB,39}\Delta t_{\rm FRB,-3}^{-4}P_{\rm day}^{-4/3}R_{0,\odot}^{3}.\nonumber\\
\ee
The observed flux at optical band is estimated by
\be
F_\nu&\simeq&\frac{2\pi\nu^2}{c^2}kT_{\rm eff}\left(\frac{R}{d}\right)^2\nonumber\\
&=&5.4\times10^{-3}~{\rm Jy}\eta f_{b,-1}^{-1}\Delta t_{\rm FRB,-3}R_{0,\odot}^{5/4}\lambda_{\rm opt}^{-2}d_{\rm 1,kpc}^{-2}.\label{flux1b}
\ee

\subsection{Case B: companion star filling Roche lobe}\label{sec22}

\begin{figure*}[]
    \centering
	\includegraphics[width = 0.49\linewidth]{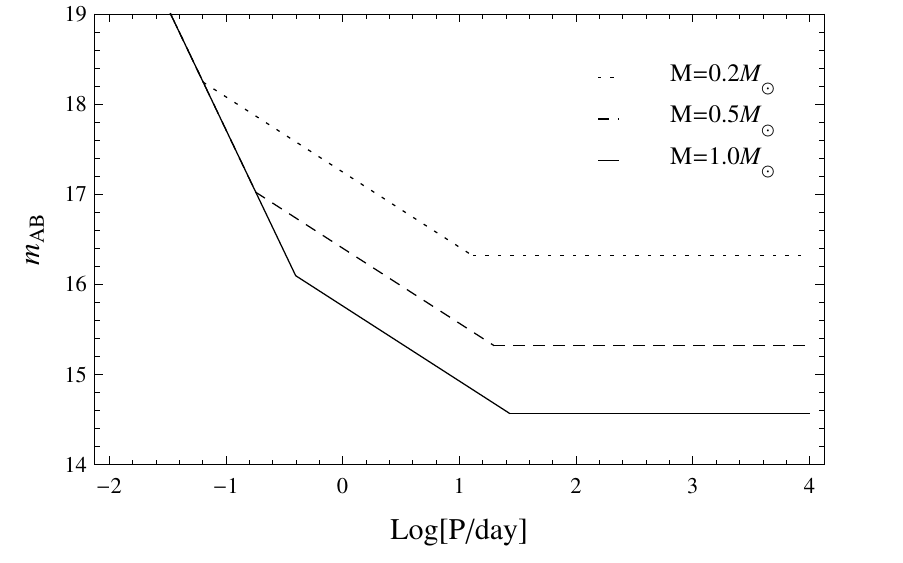}
	\includegraphics[width = 0.49\linewidth]{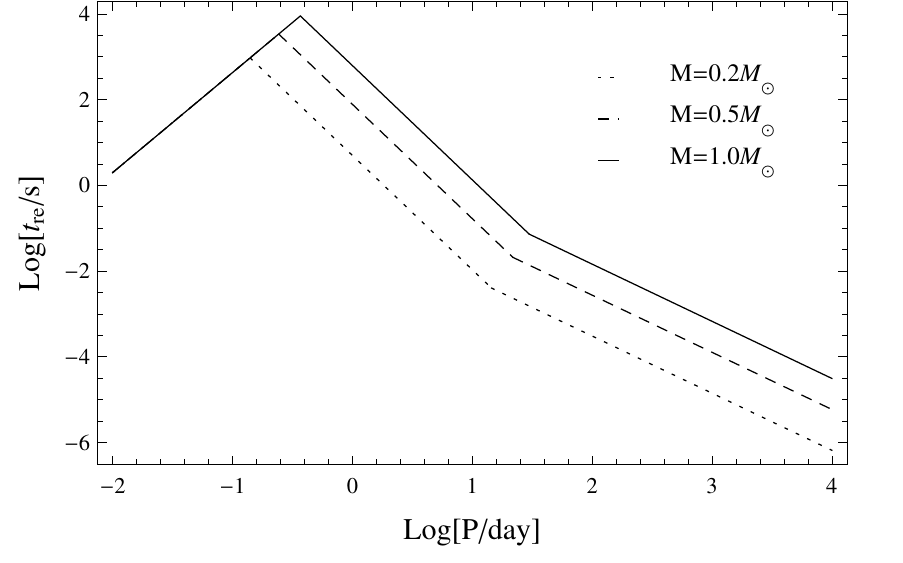}
    \caption{Left panel: the relation between optical magnitude of the re-emission and binary orbital period. Right panel: the relation between re-emission duration and binary orbital period. The dotted, dashed and solid lines represent a companion star with initial mass $M=0.2~M_\odot,0.5~M_\odot,1~M_\odot$, respectively. The FRB energy is taken as $E_{\rm FRB}=10^{39}~{\rm erg}$, and the source distance is taken as $d=10~{\rm kpc}$.}\label{fig2} 
\end{figure*} 

Next, we consider that the companion star in the binary system has filled its Roche lobe, in which case the mass of the companion star would depend on the orbital semimajor or period.
The Roche radius of the companion with mass $M$ is approximately given by \citep[e.g.,][]{Frank02}
\be
R_{\rm Roche}=0.462~a\left(\frac{M}{M_{\rm tot}}\right)^{1/3}\label{Roche}
\ee
as an approximation for the range of $0.1\lesssim M/M_{\rm NS}\lesssim 0.8$.
For a companion star filling the Roche lobe, its radius satisfies $R\simeq R_{\rm Roche}$. According to Eq.(\ref{K3}) and Eq.(\ref{Roche}), the radius-mass is approximately given by 
\be
M\simeq M_{\odot}\left(\frac{R}{R_{\odot}}\right)=2.7M_{\odot}P_{\rm day}.\label{fillingmass}
\ee
One can see that the mass of the companion star filling its Roche lobe only depends on the orbital period or separation.
For the companion star with mass $M<1M_{\odot}$, the orbital period is required to satisfies $P<0.4~{\rm day}$, if its Roche lobe has been filling.
In this case, the solid angle of the companion star opened to the FRB source is 
\be
\Delta\Omega\sim\frac{\pi R_{\rm Roche}^2}{a^2}=0.84~P_{\rm day}^{2/3}\label{Omega2}
\ee
The following discussion is similar to the case of the companion star not filling the Roche lobe discussed in Section \ref{sec21}.
According to Eq. (\ref{Omega2}), the FRB radiation energy toward to the companion star is
\be
E_{\Delta\Omega}\simeq\left(\frac{\Delta\Omega}{4\pi}\right)E_{\rm FRB}\sim7\times10^{37}~{\rm erg}~E_{\rm FRB,39}P_{\rm day}^{2/3}.\nonumber\\
\ee 
Based on Eq.(\ref{mass}) and Eq.(\ref{fillingmass}), the mass of shocked medium in this case is
\be
m(<P_{\rm FRB})
=4.5\times10^{-7}M_{\odot}E_{\rm FRB,39}\Delta t_{\rm FRB,-3}^{-1}P_{\rm day}^{5/3},\nonumber\\
\ee
and the number of shocked particles is given by
\be
N&\simeq&\frac{f_bm(<P_{\rm FRB})}{m_p}=5.4\times10^{49}f_{b,-1}\nonumber\\
&\times&E_{\rm FRB,39}\Delta t_{\rm FRB,-3}^{-1}P_{\rm day}^{5/3}.
\ee
The thermal energy assigned to each particles by thermalization process is
\be
kT\simeq\frac{\eta E_{\Delta\Omega}}{N}=0.8~{\rm eV}\eta f_{b,-1}^{-1}\Delta t_{\rm FRB,-3}P_{\rm day}^{-1},
\ee
and the mass density of the shocked medium is
\be
\rho&\simeq&\frac{m_p P_{\rm FRB}}{kT}=2.6\times10^{-5}~{\rm g~cm^{-3}}~\eta^{-1} f_{b,-1}\nonumber\\
&\times&E_{\rm FRB,39}\Delta t_{\rm FRB,-3}^{-2}P_{\rm day}^{-1/3}.\label{rho2}
\ee
The thickness of shocked medium is estimated by
\be
l\simeq\frac{f_bm(<P_{\rm FRB})}{2\pi R^2 \rho}=1.5\times10^7~{\rm cm}~\eta\Delta t_{\rm FRB,-3}.
\ee
Therefore, the optical depth for Thomson scattering is 
\be
\tau\simeq\kappa\rho l=160~f_{b,-1}E_{\rm FRB,39}\Delta t_{\rm FRB,-3}^{-1}P_{\rm day}^{-1/3}.\nonumber\\
\ee
For the orbital period of $P<0.4~{\rm day}$ we are interested here, the optical depth always satisfies $\tau\gg1$, then the effective temperature is
\be
kT_{\rm eff}&\simeq& kT\left(\frac{3}{4}\tau\right)^{-1/4}=0.24~{\rm eV}~\eta f_{b,-1}^{-5/4}\nonumber\\
&\times&E_{\rm FRB,39}^{-1/4}\Delta t_{\rm FRB,-3}^{5/4}P_{\rm day}^{-11/12}.
\ee
The re-emission luminosity from the companion star heated by an FRB is
\be
L_{\rm re}&\simeq&2\pi R^2\sigma T_{\rm eff}^4=7.6\times10^{32}~{\rm erg~s^{-1}}\eta^4 f_{b,-1}^{-5}\nonumber\\
&\times&E_{\rm FRB,39}^{-1}\Delta t_{\rm FRB,-3}^5P_{\rm day}^{-5/3},
\ee
and the corresponding typical timescale of the re-emission is about
\be
t_{\rm re}\simeq\frac{\eta E_{\Delta\Omega}}{L_{\rm re}}=9.2\times10^{4}~{\rm s}~\eta^{-3} f_{b,-1}^{5}E_{\rm FRB,39}^2\Delta t_{\rm FRB,-3}^{-5}P_{\rm day}^{7/3}.\nonumber\\
\ee
For an optical band at $\lambda\simeq5000 \text{\AA}$, the observed flux is given by
\be
F_\nu&\simeq&\frac{2\pi\nu^2}{c^2}kT_{\rm eff}\left(\frac{R}{d}\right)^2=3.6\times10^{-3}~{\rm Jy}\eta f_{b,-1}^{-5/4}\nonumber\\
&\times&E_{\rm FRB,39}^{-1/4}\Delta t_{\rm FRB,-3}^{5/4}P_{\rm day}^{13/12}\lambda_{\rm opt}^{-2}d_{\rm 1,kpc}^{-2}.\nonumber\\\label{flux2}
\ee 

Finally, we predict the spectrum and lightcurve of the re-emission. Since the re-emission is blackbody radiation, its spectrum satisfies Planck law. For a given optical band, the observed flux is proportional to the effective temperature, $F_\nu\propto T_{\rm eff}$. Thus, the lightcurve would mainly depend on the temperature evolution of the heated surface of the companion star. The internal energy stored by shocked particles is $U\simeq\eta E_{\Delta\Omega}\sim NkT$, and the temperature evolution is given by $dU/dt\simeq-2\pi R^2\sigma T_{\rm eff}^4$. Taking $T_{\rm eff}\sim T$ as an approximation, the cooling time from temperature $T_0$ to $T$  could be approximately solved as
\be
t-t_0=\frac{Nk}{6\pi R^2\sigma}\left(\frac{1}{T^3}-\frac{1}{T_0^3}\right)
\ee
where $T_0$ is the initial temperature at time $t_0$. For $T\ll T_0$, the temperature evolves as $T\propto (t/t_{\rm re})^{-1/3}$. Therefore, the optical flux decreases as $F_\nu\propto T\propto (t/t_{\rm re})^{-1/3}$, then trends to the constant optical flux of the companion star without heating effect.

\section{Result and Discussion}\label{sec3}

Based on Eq.(\ref{flux1a}), Eq.(\ref{flux1b}) and Eq.(\ref{flux2}), we plot the results as shown in Figure \ref{fig2} taking the source distance as $d\simeq10~{\rm kpc}$, the FRB isotropic energy as $E_{\rm FRB}=10^{39}~{\rm erg}$, and the FRB duration as $\Delta t_{\rm FRB}=1~{\rm ms}$. The optical magnitude is calculated by $m_{\rm AB}=-2.5\log(F_\nu/3631~{\rm Jy})$.
For the case that the companion star not filling its Roche lobe, the larger the companion star mass, the brighter the re-emission and the longer the duration. For a given companion star mass, the larger the separation the shorter the duration and the brighter the re-emission. The brightness of the re-emission trend to constant values as the separation increase.
On the other hand, for the case that the companion star filling its Roche lobe, the brightness and duration only depend on the separation. The larger the separation, the brighter the re-emission and the longer the duration.
If the binary system at distance $d\sim10~{\rm kpc}$ has the orbital period of $P\sim1~{\rm day}$ and the companion star mass of $M\sim M_{\odot}$, the optical magnitude of the re-emission would be $m_{\rm AB}\sim15.8~{\rm mag}$ (the absolute magnitude is $m_{\rm abs}\sim0.8~{\rm mag}$), and its duration is $t_{\rm re}\sim10^3~{\rm s}$.
For an optical telescope by typical optical transient survey with limiting magnitude $m_{\rm lim}=20~{\rm mag}$, the threshold flux is $F_{\rm \nu,th}=3.6\times10^{-5}~{\rm Jy}$, and the threshold distance is about $d\sim71~{\rm kpc}$. 
For a large aperture telescope with limiting magnitude $m_{\rm lim}=25~{\rm mag}$, the threshold flux is $F_{\rm \nu,th}=3.6\times10^{-7}~{\rm Jy}$, and the threshold distance is about $d\sim0.7~{\rm Mpc}$. FRB 20200120E was found in the M81 galaxy at $d=3.6~{\rm Mpc}$ \citep{Bhardwaj21,Kirsten21}. 
If a bright radio burst with energy compared with extragalactic FRBs was emitted from this source, the apparent magnitude from the heated companion star would be $m_{\rm AB}=28.5~{\rm mag}$.
Therefore, such an optical transient would be more likely detected from Galactic FRB sources.
Recently, some optical follow-up observations were performed to detect the optical counterparts of FRBs \citep{Tominaga18,Tingay19,Marnoch20,Tingay20,Kilpatrick21,Nunez21,Xin21}. Since the predict duration in this scenario is less then a few times $10^4$ seconds, a short exposure time is preferred to search and identify the optical transient. For example, \citet{Kilpatrick21} gave a limiting magnitude $m_{\rm lim}=24~{\rm mag}$ with exposure $30~{\rm s}$, \citet{Tingay19} gave a limiting magnitude $m_{\rm lim}=16~{\rm mag}$ with exposure $30~{\rm min}$, and \citet{Xin21} gave a limiting magnitude $m_{\rm lim}=15.4~{\rm mag}$ with exposure $10~{\rm s}$.
These capabilities are helpful to search Galactic optical transients by FRBs heating companion stars.
On the other hand, considering that such a transient might be in a globular cluster, one need judge whether it is resolvable.
The number density of main sequence stars in a globular cluster similar to the host of FRB 20200120E is $n\sim10^5~{\rm pc^{-3}}$ at its center. The separation is $l\sim n^{-1/3}\sim0.02~{\rm pc}~n_{\rm 5,pc^{-3}}^{-1/3}$, and the angle separation is $l/d=0.4''~d_{\rm 1,kpc}^{-1}n_{\rm 5,pc^{-3}}^{-1/3}$, which is resolvable for an optical telescope with sub-arcsec resolution. 

In the above discussion, we assume that the FRB beaming solid angle is larger than the solid angle that the companion star opened to the FRB source, as shown in Figure \ref{fig1}. In this case, if the FRB beaming direction points to the observer, we can see both the FRB and its re-emission. However, if the FRB beaming solid angle is smaller than the solid angle that the companion star opened to the FRB source, or the FRB hits the companion star but not points to the observer, one may just the an optical flare. Therefore, we propose that some optical flares in Galactic globular clusters might be triggered by FRBs generated by active neutron stars, which might appear observable repeating radio behavior in the future. On the other hand, if the FRB misses the star, the heating effect would not be produced. We assume that the intrinsic beaming solid angle of an FRB is $\Delta\Omega_{\rm FRB}$. 
For a binary system with an FRB source, the solid angle of the companion star opened to the FRB source is $\Delta \Omega$ given by Eq.(\ref{Omega}) and Eq.(\ref{Omega2}). If the radiation beam is random, the probability that the FRB hits the companion star would be $p\sim(\Delta\Omega+\Delta\Omega_{\rm FRB})/4\pi$. If $\Delta\Omega_{\rm FRB}\ll\Delta\Omega$, one has $p\sim0.01R_{0,\odot}^2P_{\rm day}^{-4/3}$ for the companion star not filling its Roche lobe and $p\sim0.07P_{\rm day}^{2/3}$ for the companion star filling its Roche lobe; if $\Delta\Omega_{\rm FRB}\gg\Delta\Omega$, one has $p\sim\Delta\Omega_{\rm FRB}/4\pi$.

The periodic activities of some FRB sources implies that they might be in binary systems.
Besides FRB 180916B with 16.35-day periodic activity \citep{CHIME20b}, some FRB sources and Galactic magnetars also appear possible periodic activities. The first repeating source, FRB 121102, showed a possible long period of $\sim160~{\rm day}$ \citep{Rajwade20}, and the X-ray bursts from Galactic magnetars SGR 1806-20 and SGR 1935+2154 were also found to have periodic activities of 398.2 day \citep{Zhang21b} and 237 day \citep{Zou21}, respectively. If these periodic activities are indeed due to the orbital periods of binary systems, FRB sources with bright radio bursts similar to the typical extragalactic FRBs might have opportunity to produce the observable optical transients if they are in Milky Way. 

In this work, we mainly focus on the interaction process between the strong radio emission of an FRB and its companion star, and predict its multiwavelength observation. If the radio emission energy is only a small fraction of the total energy required by the FRB central engine, the heating effect would be more significant than that we estimate here. 
The above results would replace $E_{\rm FRB}$ with $E_{\rm FRB}/\xi$, where $\xi$ is the radio emission fraction. For example, the isotropic energy of Galactic FRB 200428 is $E_{\rm FRB}\sim10^{35}~{\rm erg}$, and the energy of the associated X-ray burst is $E_{\rm XRB}\sim10^{40}~{\rm erg}$. 
Thus, the radio emission fraction would be $\xi\sim10^{-5}$. If such an FRB source with $E_{\rm FRB}\sim10^{35}~{\rm erg}$ and $\xi\sim10^{-5}$ is in a close binary system in the Milky Way, the optical brightness by the heating effect would be comparable with or even larger the typical value given by the above discussion. However, on the other hand, the large burst event rate of FRB 121102 suggested that its radio emission fraction can not be as small as that of FRB 200428 \citep{Yang21}. 
Based on the observed event rate of FRB 121102 \citep{Li21}, if its central engine is a magnetar with magnetic field of $\sim10^{15}~{\rm G}$ and has the radio emission fraction as same as that of FRB 200428, its active age would be less than one year, which is much less than the observed active age of $\gtrsim8~{\rm yr}$. This result implies that FRB 121102 has a much larger radio emission fraction than that of FRB 200428.

\acknowledgments

We thank the anonymous referee for constructive comments and suggestions, which clarified many concepts and improved the paper.
We also thank Bing Zhang for reading the manuscript and for his excellent comments. This work is supported by National Natural Science Foundation of China grant No. 12003028.


\begin{thebibliography}{}
\expandafter\ifx\csname natexlab\endcsname\relax\def\natexlab#1{#1}\fi

\bibitem[{{Beloborodov}(2017)}]{Beloborodov17}
{Beloborodov}, A.~M. 2017, \apjl, 843, L26

\bibitem[{{Bhardwaj} {et~al.}(2021){Bhardwaj}, {Gaensler}, {Kaspi},
  {Landecker}, {Mckinven}, {Michilli}, {Pleunis}, {Tendulkar}, {Andersen},
  {Boyle}, {Cassanelli}, {Chawla}, {Cook}, {Dobbs}, {Fonseca}, {Kaczmarek},
  {Leung}, {Masui}, {Mnchmeyer}, {Ng}, {Rafiei-Ravandi}, {Scholz}, {Shin},
  {Smith}, {Stairs}, \& {Zwaniga}}]{Bhardwaj21}
{Bhardwaj}, M., {Gaensler}, B.~M., {Kaspi}, V.~M., {et~al.} 2021, \apjl, 910,
  L18

\bibitem[{{Bochenek} {et~al.}(2020){Bochenek}, {Ravi}, {Belov}, {Hallinan},
  {Kocz}, {Kulkarni}, \& {McKenna}}]{Bochenek20}
{Bochenek}, C.~D., {Ravi}, V., {Belov}, K.~V., {et~al.} 2020, \nat, 587, 59

\bibitem[{{CHIME/FRB Collaboration} {et~al.}(2020{\natexlab{a}}){CHIME/FRB
  Collaboration}, {Andersen}, {Bandura}, {Bhardwaj}, {Bij}, {Boyce}, {Boyle},
  {Brar}, {Cassanelli}, {Chawla}, {Chen}, {Cliche}, {Cook}, {Cubranic},
  {Curtin}, {Denman}, {Dobbs}, {Dong}, {Fandino}, {Fonseca}, {Gaensler},
  {Giri}, {Good}, {Halpern}, {Hill}, {Hinshaw}, {H{\"o}fer}, {Josephy},
  {Kania}, {Kaspi}, {Landecker}, {Leung}, {Li}, {Lin}, {Masui}, {McKinven},
  {Mena-Parra}, {Merryfield}, {Meyers}, {Michilli}, {Milutinovic},
  {Mirhosseini}, {M{\"u}nchmeyer}, {Naidu}, {Newburgh}, {Ng}, {Patel}, {Pen},
  {Pinsonneault-Marotte}, {Pleunis}, {Quine}, {Rafiei-Ravandi}, {Rahman},
  {Ransom}, {Renard}, {Sanghavi}, {Scholz}, {Shaw}, {Shin}, {Siegel}, {Singh},
  {Smegal}, {Smith}, {Stairs}, {Tan}, {Tendulkar}, {Tretyakov}, {Vanderlinde},
  {Wang}, {Wulf}, \& {Zwaniga}}]{CHIME20}
{CHIME/FRB Collaboration}, {Andersen}, B.~C., {Bandura}, K.~M., {et~al.}
  2020{\natexlab{a}}, \nat, 587, 54

\bibitem[{{CHIME/FRB Collaboration} {et~al.}(2020{\natexlab{b}}){CHIME/FRB
  Collaboration}, {Amiri}, {Andersen}, {Bandura}, {Bhardwaj}, {Boyle}, {Brar},
  {Chawla}, {Chen}, {Cliche}, {Cubranic}, {Deng}, {Denman}, {Dobbs}, {Dong},
  {Fandino}, {Fonseca}, {Gaensler}, {Giri}, {Good}, {Halpern}, {Hessels},
  {Hill}, {H{\"o}fer}, {Josephy}, {Kania}, {Karuppusamy}, {Kaspi}, {Keimpema},
  {Kirsten}, {Landecker}, {Lang}, {Leung}, {Li}, {Lin}, {Marcote}, {Masui},
  {McKinven}, {Mena-Parra}, {Merryfield}, {Michilli}, {Milutinovic},
  {Mirhosseini}, {Naidu}, {Newburgh}, {Ng}, {Nimmo}, {Paragi}, {Patel}, {Pen},
  {Pinsonneault-Marotte}, {Pleunis}, {Rafiei-Ravandi}, {Rahman}, {Ransom},
  {Renard}, {Sanghavi}, {Scholz}, {Shaw}, {Shin}, {Siegel}, {Singh}, {Smegal},
  {Smith}, {Stairs}, {Tendulkar}, {Tretyakov}, {Vanderlinde}, {Wang}, {Wang},
  {Wulf}, {Yadav}, \& {Zwaniga}}]{CHIME20b}
{CHIME/FRB Collaboration}, {Amiri}, M., {Andersen}, B.~C., {et~al.}
  2020{\natexlab{b}}, \nat, 582, 351

\bibitem[Connor et al.(2020)]{Connor20} Connor, L., Miller, M.~C., \& Gardenier, D.~W.\ 2020, \mnras, 497, 3076

\bibitem[{{Cooray} {et~al.}(2012){Cooray}, {Gong}, {Smidt}, \&
  {Santos}}]{Cooray}
{Cooray}, A., {Gong}, Y., {Smidt}, J., \& {Santos}, M.~G. 2012, \apj, 756, 92

\bibitem[{{Cordes} \& {Chatterjee}(2019)}]{Cordes19}
{Cordes}, J.~M., \& {Chatterjee}, S. 2019, \araa, 57, 417

\bibitem[{{Dai} \& {Zhong}(2020)}]{Dai20}
{Dai}, Z.~G., \& {Zhong}, S.~Q. 2020, \apjl, 895, L1

\bibitem[{{Deng} {et~al.}(2021){Deng}, {Zhong}, \& {Dai}}]{Deng21}
{Deng}, C.-M., {Zhong}, S.-Q., \& {Dai}, Z.-G. 2021, arXiv e-prints,
  arXiv:2102.06796

\bibitem[{{Frank} {et~al.}(2002){Frank}, {King}, \& {Raine}}]{Frank02}
{Frank}, J., {King}, A., \& {Raine}, D.~J. 2002, {Accretion Power in
  Astrophysics: Third Edition}

\bibitem[{{Geng} {et~al.}(2021){Geng}, {Li}, \& {Huang}}]{Geng21}
{Geng}, J.-J., {Li}, B., \& {Huang}, Y.-F. 2021, arXiv e-prints,
  arXiv:2103.04165

\bibitem[{{Guenther} {et~al.}(1992){Guenther}, {Demarque}, {Kim}, \&
  {Pinsonneault}}]{Guenther92}
{Guenther}, D.~B., {Demarque}, P., {Kim}, Y.~C., \& {Pinsonneault}, M.~H. 1992,
  \apj, 387, 372

\bibitem[{{Ioka}(2020)}]{Ioka20}
{Ioka}, K. 2020, \apjl, 904, L15

\bibitem[{{Ioka} \& {Zhang}(2020)}]{Ioka20b}
{Ioka}, K., \& {Zhang}, B. 2020, \apjl, 893, L26

\bibitem[{{Katz}(2016)}]{Katz16}
{Katz}, J.~I. 2016, \apj, 826, 226

\bibitem[Kilpatrick et al.(2021)]{Kilpatrick21} Kilpatrick, C.~D., Burchett, J.~N., Jones, D.~O., et al.\ 2021, \apjl, 907, L3

\bibitem[{{Kippenhahn} {et~al.}(2012){Kippenhahn}, {Weigert}, \&
  {Weiss}}]{Kippenhahn12}
{Kippenhahn}, R., {Weigert}, A., \& {Weiss}, A. 2012, {Stellar Structure and
  Evolution}

\bibitem[{{Kirsten} {et~al.}(2021){Kirsten}, {Marcote}, {Nimmo}, {Hessels},
  {Bhardwaj}, {Tendulkar}, {Keimpema}, {Yang}, {Snelders}, {Scholz},
  {Pearlman}, {Law}, {Peters}, {Giroletti}, {Hewitt}, {Bach}, {Bezukovs},
  {Burgay}, {Buttaccio}, {Conway}, {Corongiu}, {Feiler}, {Forss{\'e}n},
  {Gawro{\'n}ski}, {Karuppusamy}, {Kharinov}, {Lindqvist}, {Maccaferri},
  {Melnikov}, {Ould-Boukattine}, {Paragi}, {Possenti}, {Surcis}, {Wang},
  {Yuan}, {Aggarwal}, {Anna-Thomas}, {Bower}, {Blaauw}, {Burke-Spolaor},
  {Cassanelli}, {Clarke}, {Fonseca}, {Gaensler}, {Gopinath}, {Kaspi}, {Kassim},
  {Lazio}, {Leung}, {Li}, {Lin}, {Masui}, {Mckinven}, {Michilli}, {Mikhailov},
  {Ng}, {Orbidans}, {Pen}, {Petroff}, {Rahman}, {Ransom}, {Shin}, {Smith},
  {Stairs}, \& {Vlemmings}}]{Kirsten21}
{Kirsten}, F., {Marcote}, B., {Nimmo}, K., {et~al.} 2021, arXiv e-prints,
  arXiv:2105.11445

\bibitem[{{Kremer} {et~al.}(2021){Kremer}, {Piro}, \& {Li}}]{Kremer21}
{Kremer}, K., {Piro}, A.~L., \& {Li}, D. 2021, arXiv e-prints, arXiv:2107.03394

\bibitem[{{Kumar} \& {Bo{\v{s}}njak}(2020)}]{Kumar20}
{Kumar}, P., \& {Bo{\v{s}}njak}, {\v{Z}}. 2020, \mnras, 494, 2385

\bibitem[{{Kumar} {et~al.}(2017){Kumar}, {Lu}, \& {Bhattacharya}}]{Kumar17}
{Kumar}, P., {Lu}, W., \& {Bhattacharya}, M. 2017, \mnras, 468, 2726

\bibitem[{{Kundu} \& {Zhang}(2021)}]{Kundu21}
{Kundu}, E., \& {Zhang}, B. 2021, arXiv e-prints, arXiv:2107.12989

\bibitem[{{Lejeune} \& {Schaerer}(2001)}]{Lejeune01}
{Lejeune}, T., \& {Schaerer}, D. 2001, \aap, 366, 538

\bibitem[{{Li} {et~al.}(2020){Li}, {Lin}, {Xiong}, {Ge}, {Li}, {Li}, {Lu},
  {Zhang}, {Tuo}, {Nang}, {Zhang}, {Xiao}, {Chen}, {Song}, {Xu}, {Liu}, {Jia},
  {Cao}, {Qu}, {Zhang}, {Gu}, {Liao}, {Zhao}, {Tan}, {Nie}, {Zhao}, {Zheng},
  {Zheng}, {Luo}, {Cai}, {Li}, {Xue}, {Bu}, {Chang}, {Chen}, {Chen}, {Chen},
  {Chen}, {Cui}, {Cui}, {Deng}, {Dong}, {Du}, {Fu}, {Gao}, {Gao}, {Gao}, {Gu},
  {Guan}, {Guo}, {Han}, {Huang}, {Huo}, {Jiang}, {Jiang}, {Jin}, {Jin}, {Kong},
  {Li}, {Li}, {Li}, {Li}, {Li}, {Li}, {Li}, {Liang}, {Liu}, {Liu}, {Liu},
  {Liu}, {Liu}, {Lu}, {Lu}, {Luo}, {Ma}, {Meng}, {Ou}, {Sai}, {Shang}, {Song},
  {Sun}, {Tao}, {Wang}, {Wang}, {Wang}, {Wang}, {Wang}, {Wen}, {Wu}, {Wu},
  {Wu}, {Xiao}, {Xu}, {Yang}, {Yang}, {Yang}, {Yang}, {Yi}, {Yin}, {You},
  {Zhang}, {Zhang}, {Zhang}, {Zhang}, {Zhang}, {Zhang}, {Zhang}, {Zhang},
  {Zhang}, {Zhang}, {Zhang}, {Zhang}, {Zhang}, {Zhang}, {Zhang}, {Zhang},
  {Zhou}, {Zhou}, {Zhu}, {Zhu}, \& {Zhuang}}]{Li20}
{Li}, C.~K., {Lin}, L., {Xiong}, S.~L., {et~al.} 2020, arXiv e-prints,
  arXiv:2005.11071

\bibitem[Li et al.(2021a)]{Li21} Li, D., Wang, P., Zhu, W.~W., et al.\ 2021a, arXiv:2107.08205

\bibitem[{{Li} {et~al.}(2021b){Li}, {Yang}, {Wang}, {Xu}, {Shao}, {Liu}, \&
  {Dai}}]{Li21c}
{Li}, Q.-C., {Yang}, Y.-P., {Wang}, F.~Y., {et~al.} 2021b, arXiv e-prints,
  arXiv:2108.00350

\bibitem[{{Lorimer} {et~al.}(2007){Lorimer}, {Bailes}, {McLaughlin},
  {Narkevic}, \& {Crawford}}]{Lorimer07}
{Lorimer}, D.~R., {Bailes}, M., {McLaughlin}, M.~A., {Narkevic}, D.~J., \&
  {Crawford}, F. 2007, Science, 318, 777

\bibitem[{{Lu} {et~al.}(2021){Lu}, {Beniamini}, \& {Kumar}}]{Lu21}
{Lu}, W., {Beniamini}, P., \& {Kumar}, P. 2021, arXiv e-prints,
  arXiv:2107.04059

\bibitem[{{Lu} {et~al.}(2020){Lu}, {Kumar}, \& {Zhang}}]{Lu20}
{Lu}, W., {Kumar}, P., \& {Zhang}, B. 2020, \mnras, 498, 1397

\bibitem[Luo et al.(2020)]{Luo20} Luo, R., Men, Y., Lee, K., et al.\ 2020, \mnras, 494, 665

\bibitem[{{Lyubarsky}(2021)}]{Lyubarsky21}
{Lyubarsky}, Y. 2021, arXiv e-prints, arXiv:2103.00470

\bibitem[{{Lyutikov} {et~al.}(2020){Lyutikov}, {Barkov}, \&
  {Giannios}}]{Lyutikov20}
{Lyutikov}, M., {Barkov}, M.~V., \& {Giannios}, D. 2020, \apjl, 893, L39

\bibitem[{{Macchi} {et~al.}(2013){Macchi}, {Borghesi}, \& {Passoni}}]{Macchi13}
{Macchi}, A., {Borghesi}, M., \& {Passoni}, M. 2013, Reviews of Modern Physics,
  85, 751

\bibitem[{{Margalit} {et~al.}(2020){Margalit}, {Beniamini}, {Sridhar}, \&
  {Metzger}}]{Margalit20}
{Margalit}, B., {Beniamini}, P., {Sridhar}, N., \& {Metzger}, B.~D. 2020,
  \apjl, 899, L27

\bibitem[Marnoch et al.(2020)]{Marnoch20} Marnoch, L., Ryder, S.~D., Bannister, K.~W., et al.\ 2020, \aap, 639, A119

\bibitem[{{Mereghetti} {et~al.}(2020){Mereghetti}, {Savchenko}, {Ferrigno},
  {G{\"o}tz}, {Rigoselli}, {Tiengo}, {Bazzano}, {Bozzo}, {Coleiro},
  {Courvoisier}, {Doyle}, {Goldwurm}, {Hanlon}, {Jourdain}, {von Kienlin},
  {Lutovinov}, {Martin-Carrillo}, {Molkov}, {Natalucci}, {Onori}, {Panessa},
  {Rodi}, {Rodriguez}, {S{\'a}nchez-Fern{\'a}ndez}, {Sunyaev}, \&
  {Ubertini}}]{Mereghetti20}
{Mereghetti}, S., {Savchenko}, V., {Ferrigno}, C., {et~al.} 2020, \apjl, 898,
  L29

\bibitem[{{Metzger} {et~al.}(2019){Metzger}, {Margalit}, \&
  {Sironi}}]{Metzger19}
{Metzger}, B.~D., {Margalit}, B., \& {Sironi}, L. 2019, \mnras, 485, 4091

\bibitem[{{Murase} {et~al.}(2016){Murase}, {Kashiyama}, \&
  {M{\'e}sz{\'a}ros}}]{Murase16}
{Murase}, K., {Kashiyama}, K., \& {M{\'e}sz{\'a}ros}, P. 2016, \mnras, 461,
  1498

\bibitem[Nu{\~n}ez et al.(2021)]{Nunez21} Nu{\~n}ez, C., Tejos, N., Pignata, G., et al.\ 2021, arXiv:2104.09727


\bibitem[{{Petroff} {et~al.}(2019){Petroff}, {Hessels}, \&
  {Lorimer}}]{Petroff19}
{Petroff}, E., {Hessels}, J.~W.~T., \& {Lorimer}, D.~R. 2019, \aapr, 27, 4

\bibitem[{{Rajwade} {et~al.}(2020){Rajwade}, {Mickaliger}, {Stappers},
  {Morello}, {Agarwal}, {Bassa}, {Breton}, {Caleb}, {Karastergiou}, {Keane}, \&
  {Lorimer}}]{Rajwade20}
{Rajwade}, K.~M., {Mickaliger}, M.~B., {Stappers}, B.~W., {et~al.} 2020,
  \mnras, 495, 3551

\bibitem[{{Ridnaia} {et~al.}(2020){Ridnaia}, {Svinkin}, {Frederiks}, {Bykov},
  {Popov}, {Aptekar}, {Golenetskii}, {Lysenko}, {Tsvetkova}, {Ulanov}, \&
  {Cline}}]{Ridnaia20}
{Ridnaia}, A., {Svinkin}, D., {Frederiks}, D., {et~al.} 2020, arXiv e-prints,
  arXiv:2005.11178

\bibitem[{{Schaerer}(2002)}]{Schaerer02}
{Schaerer}, D. 2002, \aap, 382, 28

\bibitem[{{Somov}(2012)}]{Somov12}
{Somov}, B.~V. 2012, {Plasma Astrophysics, Part I}, Vol. 391

\bibitem[{{Sridhar} {et~al.}(2021){Sridhar}, {Metzger}, {Beniamini},
  {Margalit}, {Renzo}, {Sironi}, \& {Kovlakas}}]{Sridhar21}
{Sridhar}, N., {Metzger}, B.~D., {Beniamini}, P., {et~al.} 2021, arXiv
  e-prints, arXiv:2102.06138

\bibitem[{{Tavani} {et~al.}(2020){Tavani}, {Casentini}, {Ursi}, {Verrecchia},
  {Addis}, {Antonelli}, {Argan}, {Barbiellini}, {Baroncelli}, {Bernardi},
  {Bianchi}, {Bulgarelli}, {Caraveo}, {Cardillo}, {Cattaneo}, {Chen}, {Costa},
  {Del Monte}, {Di Cocco}, {Di Persio}, {Donnarumma}, {Evangelista}, {Feroci},
  {Ferrari}, {Fioretti}, {Fuschino}, {Galli}, {Gianotti}, {Giuliani},
  {Labanti}, {Lazzarotto}, {Lipari}, {Longo}, {Lucarelli}, {Magro},
  {Marisaldi}, {Mereghetti}, {Morelli}, {Morselli}, {Naldi}, {Pacciani},
  {Parmiggiani}, {Paoletti}, {Pellizzoni}, {Perri}, {Perotti}, {Piano},
  {Picozza}, {Pilia}, {Pittori}, {Puccetti}, {Pupillo}, {Rapisarda},
  {Rappoldi}, {Rubini}, {Setti}, {Soffitta}, {Trifoglio}, {Trois},
  {Vercellone}, {Vittorini}, {Giommi}, \& {D' Amico}}]{Tavani20}
{Tavani}, M., {Casentini}, C., {Ursi}, A., {et~al.} 2020, arXiv e-prints,
  arXiv:2005.12164

\bibitem[{{Tendulkar} {et~al.}(2021){Tendulkar}, {Gil de Paz}, {Kirichenko},
  {Hessels}, {Bhardwaj}, {{\'A}vila}, {Bassa}, {Chawla}, {Fonseca}, {Kaspi},
  {Keimpema}, {Kirsten}, {Lazio}, {Marcote}, {Masui}, {Nimmo}, {Paragi},
  {Rahman}, {Pay{\'a}}, {Scholz}, \& {Stairs}}]{Tendulkar21}
{Tendulkar}, S.~P., {Gil de Paz}, A., {Kirichenko}, A.~Y., {et~al.} 2021,
  \apjl, 908, L12

\bibitem[{{The CHIME/FRB Collaboration} {et~al.}(2021){The CHIME/FRB
  Collaboration}, {:}, {Amiri}, {Andersen}, {Bandura}, {Berger}, {Bhardwaj},
  {Boyce}, {Boyle}, {Brar}, {Breitman}, {Cassanelli}, {Chawla}, {Chen},
  {Cliche}, {Cook}, {Cubranic}, {Curtin}, {Deng}, {Dobbs}, {Fengqiu}, {Dong},
  {Eadie}, {Fandino}, {Fonseca}, {Gaensler}, {Giri}, {Good}, {Halpern}, {Hill},
  {Hinshaw}, {Josephy}, {Kaczmarek}, {Kader}, {Kania}, {Kaspi}, {Landecker},
  {Lang}, {Leung}, {Li}, {Lin}, {Masui}, {Mckinven}, {Mena-Parra},
  {Merryfield}, {Meyers}, {Michilli}, {Milutinovic}, {Mirhosseini},
  {M{\"u}nchmeyer}, {Naidu}, {Newburgh}, {Ng}, {Patel}, {Pen}, {Petroff},
  {Pinsonneault-Marotte}, {Pleunis}, {Rafiei-Ravandi}, {Rahman}, {Ransom},
  {Renard}, {Sanghavi}, {Scholz}, {Shaw}, {Shin}, {Siegel}, {Sikora}, {Singh},
  {Smith}, {Stairs}, {Tan}, {Tendulkar}, {Vanderlinde}, {Wang}, {Wulf}, \&
  {Zwaniga}}]{CHIME21}
{The CHIME/FRB Collaboration}, {:}, {Amiri}, M., {et~al.} 2021, arXiv e-prints,
  arXiv:2106.04352

\bibitem[{{Thornton} {et~al.}(2013){Thornton}, {Stappers}, {Bailes},
  {Barsdell}, {Bates}, {Bhat}, {Burgay}, {Burke-Spolaor}, {Champion}, {Coster},
  {D'Amico}, {Jameson}, {Johnston}, {Keith}, {Kramer}, {Levin}, {Milia}, {Ng},
  {Possenti}, \& {van Straten}}]{Thornton13}
{Thornton}, D., {Stappers}, B., {Bailes}, M., {et~al.} 2013, Science, 341, 53

\bibitem[{{Tingay} \& {Yang}(2019)}]{Tingay19}
{Tingay}, S.~J., \& {Yang}, Y.-P. 2019, \apj, 881, 30

\bibitem[Tingay(2020)]{Tingay20} Tingay, S.\ 2020, \pasa, 37, e015

\bibitem[Tominaga et al.(2018)]{Tominaga18} Tominaga, N., Niino, Y., Totani, T., et al.\ 2018, \pasj, 70, 103

\bibitem[{{Wada} {et~al.}(2021){Wada}, {Ioka}, \& {Zhang}}]{Wada21}
{Wada}, T., {Ioka}, K., \& {Zhang}, B. 2021, arXiv e-prints, arXiv:2105.14480

\bibitem[{{Wadiasingh} {et~al.}(2020){Wadiasingh}, {Beniamini}, {Timokhin},
  {Baring}, {van der Horst}, {Harding}, \& {Kazanas}}]{Wadiasingh20}
{Wadiasingh}, Z., {Beniamini}, P., {Timokhin}, A., {et~al.} 2020, \apj, 891, 82

\bibitem[{{Wang} {et~al.}(2020{\natexlab{a}}){Wang}, {Wang}, {Yang}, {Yu},
  {Zuo}, \& {Dai}}]{Wang20e}
{Wang}, F.~Y., {Wang}, Y.~Y., {Yang}, Y.-P., {et~al.} 2020{\natexlab{a}}, \apj,
  891, 72

\bibitem[{{Wang} \& {Lai}(2020)}]{Wang20c}
{Wang}, J.-S., \& {Lai}, D. 2020, \apj, 892, 135

\bibitem[{{Wang} {et~al.}(2021){Wang}, {Xu}, {Wang}, {Du}, {Cheng}, {Zheng}, \&
  {Xu}}]{Wang21}
{Wang}, W.-H., {Xu}, H., {Wang}, W.-Y., {et~al.} 2021, arXiv e-prints,
  arXiv:2107.13725

\bibitem[{{Wang} {et~al.}(2020{\natexlab{b}}){Wang}, {Li}, {Yang}, {Luo},
  {Zhang}, {Lin}, {Liang}, \& {Qin}}]{Wang20d}
{Wang}, X.-G., {Li}, L., {Yang}, Y.-P., {et~al.} 2020{\natexlab{b}}, \apjl,
  894, L22

\bibitem[{{Xiao} {et~al.}(2021){Xiao}, {Wang}, \& {Dai}}]{Xiao21}
{Xiao}, D., {Wang}, F., \& {Dai}, Z. 2021, arXiv e-prints, arXiv:2101.04907

\bibitem[Xin et al.(2021)]{Xin21} Xin, L.~P., Li, H.~L., Wang, J., et al.\ 2021, arXiv:2108.06931

\bibitem[{{Yang} \& {Zhang}(2018)}]{Yang18}
{Yang}, Y.-P., \& {Zhang}, B. 2018, \apj, 868, 31

\bibitem[{{Yang} \& {Zhang}(2020)}]{Yang20b}
---. 2020, \apjl, 892, L10

\bibitem[{{Yang} \& {Zhang}(2021)}]{Yang21}
---. 2021, arXiv e-prints, arXiv:2104.01925

\bibitem[{{Yang} {et~al.}(2016){Yang}, {Zhang}, \& {Dai}}]{Yang16}
{Yang}, Y.-P., {Zhang}, B., \& {Dai}, Z.-G. 2016, \apjl, 819, L12

\bibitem[{{Yang} {et~al.}(2019){Yang}, {Zhang}, \& {Wei}}]{Yang19}
{Yang}, Y.-P., {Zhang}, B., \& {Wei}, J.-Y. 2019, \apj, 878, 89

\bibitem[{{Yang} {et~al.}(2020){Yang}, {Zhu}, {Zhang}, \& {Wu}}]{Yang20}
{Yang}, Y.-P., {Zhu}, J.-P., {Zhang}, B., \& {Wu}, X.-F. 2020, \apjl, 901, L13

\bibitem[{{Yi} {et~al.}(2014){Yi}, {Gao}, \& {Zhang}}]{Yi14}
{Yi}, S.-X., {Gao}, H., \& {Zhang}, B. 2014, \apjl, 792, L21

\bibitem[{{Zhang}(2017)}]{Zhang17}
{Zhang}, B. 2017, \apjl, 836, L32

\bibitem[{{Zhang}(2020{\natexlab{a}})}]{Zhang20c}
---. 2020{\natexlab{a}}, \apjl, 890, L24

\bibitem[{{Zhang}(2020{\natexlab{b}})}]{Zhang20}
---. 2020{\natexlab{b}}, \nat, 587, 45

\bibitem[{{Zhang}(2020{\natexlab{c}})}]{Zhang20d}
---. 2020{\natexlab{c}}, \nat, 582, 344

\bibitem[{{Zhang} {et~al.}(2021){Zhang}, {Tu}, \& {Wang}}]{Zhang21b}
{Zhang}, G.~Q., {Tu}, Z.-L., \& {Wang}, F.~Y. 2021, arXiv e-prints,
  arXiv:2101.07923

\bibitem[{{Zou} {et~al.}(2021){Zou}, {Zhang}, {Zhang}, {Yang}, {Wang}, \&
  {Shao}}]{Zou21}
{Zou}, J.-H., {Zhang}, B.-B., {Zhang}, G.-Q., {et~al.} 2021, arXiv e-prints,
  arXiv:2107.03800

\end{thebibliography}

\end{document}